\documentclass[12pt,a4paper]{article}
\usepackage[utf8x]{inputenc}
\usepackage[T1]{fontenc}
\usepackage{makecell}
\usepackage{bm}
\usepackage{amsmath}
\usepackage{amssymb}
\usepackage[round]{natbib}
\usepackage[dvips]{epsfig}
\usepackage{dcolumn}
\usepackage{enumerate}
\usepackage{hhline}
\usepackage{dsfont}
\usepackage{afterpage}
\usepackage{arydshln}
\usepackage{graphicx}
\usepackage{color}
\usepackage[usenames,dvipsnames]{xcolor}
\usepackage{rotating}
\usepackage[breaklinks]{hyperref}
\usepackage{xr}
\usepackage[percent]{overpic}
\usepackage{subfig}
\usepackage[]{caption}
\usepackage[ruled]{algorithm2e}
\usepackage{diagbox}
\usepackage{graphicx}
\usepackage{wrapfig}
\usepackage{lscape}
\usepackage{mathabx}
\usepackage{adjustbox}
\input 
\usepackage{fontenc}
\usepackage{setspace}
\usepackage{bm}
\usepackage{slashbox}
\usepackage{lscape}
\usepackage{multirow}
\usepackage{eurosym}
\usepackage{titlesec}
\usepackage{placeins}
\usepackage[mdyy]{datetime}
\usepackage{comment}

\def\spacingset#1{\renewcommand{\baselinestretch}%
{#1}\small\normalsize} \spacingset{1}

\def \dsR {\text{$\mathds{R}$}}

\DeclareMathOperator{\diag}{diag}

\DeclareMathOperator{\ND}{\mathcal{N}}

\DeclareMathOperator{\IGD}{\mathcal{IG}}

\usepackage{color}
\usepackage{colordvi}
\newlength{\breite}
\breite\textwidth
\newcounter{aufg}[section]
  {\refstepcounter{aufg}\noindent\textbf{Exercise \arabic{aufg}:}
   \\*[1ex]\noindent}{\vspace{.5cm}}
   
 \newcounter{notes}[section]
  {\refstepcounter{aufg}\noindent\textbf{}
   \\*[1ex]\noindent}{\vspace{.5cm}}
   
\usepackage{amsthm}  

\theoremstyle{definition}

\newtheorem*{beisp*}{Example}
\newtheorem{Proof}{Proof}


\newtheoremstyle{break}
  {}
  {}
  {}
  {}
  {\bfseries}
  {.}
  {\newline}
  {}
  
\theoremstyle{break}

\begin{document} 

\begin{titlepage}

\title{Truly Multivariate Structured Additive Distributional Regression}

\author{Lucas Kock$\mbox{}^1$ and Nadja Klein$\mbox{}^{2\ast}$} 

\date{\today}
\maketitle
\thispagestyle{empty}
\noindent
\vspace{2em}

\begin{center}
{\Large Abstract}
\end{center}
\vspace{-1pt}
\noindent Generalized additive models for location, scale and shape (GAMLSS) are a popular extension to mean regression models where each parameter of an arbitrary parametric distribution is modelled through covariates. While such models have been  developed for univariate and bivariate responses, the truly multivariate case remains extremely challenging for both computational and theoretical reasons. Alternative approaches to GAMLSS may allow for higher-dimensional response vectors to be modelled jointly but often assume a fixed dependence structure not depending on covariates or are limited with respect to modelling flexibility or computational aspects. We contribute to this gap in the literature and propose a truly multivariate distributional model, which allows one to benefit from the flexibility of GAMLSS even when the response has dimension larger than two or three. Building on  copula regression, we model the dependence structure of the response through a Gaussian copula, while the marginal distributions can vary across components. Our model is highly parameterized but estimation becomes feasible with Bayesian inference employing shrinkage priors. We demonstrate the competitiveness of our approach in a simulation study and illustrate how it complements existing approaches along the examples of childhood malnutrition and a yet unexplored data set on traffic detection in Berlin.
\vspace{20pt}
 

\vspace*{\fill}
\noindent {\small\textbf{Acknowledgments:} Nadja Klein acknowledges support through the Emmy Noether grant KL 3037/1-1 of the German research foundation (DFG).}

\vspace{20pt}

\noindent{\small
$^1$ Department of Statistics and Data Science, National University of Singapore, Singapore\\
$^2$ Scientific Computing Center, Karlsruhe Institute of Technology, Germany\\
$^\ast$ Corresponding author  
}

\end{titlepage}

\spacingset{1.5} 

\section{Introduction}
Generalized additive models for location, scale, and shape \citep[GAMLSS;][]{RigSta2005}, also known as structured additive distributional regression \citep{KleKneLan2015}, extend generalized additive models  \citep[GAMs;][]{HasTib1990} towards arbitrary  parametric response distributions. While GAMs focus on modelling the conditional mean of a univariate response $Y$, GAMLSS model the entire conditional distribution by relating   each distributional parameter to covariates through flexible semiparametric predictors.

This idea can be carried over to regression data with multivariate responses. The first multivariate distributional regression model within the  framework of GAMLSS has been proposed by \citet{KleKneKlaLan2015}. These authors consider -- amongst other parameteric multivariate distributions -- the bivariate Gaussian and t-distributions, and in their model  the marginal means, standard deviations and correlation parameter (and also the degrees of freedom in case of the t-distribution) can be observation-specific. \citet{MusMaySimUmlZei2022,GioFasBrowBel2022} generalize this idea for the special case of the Gaussian distribution beyond two-dimensional responses  by parameterizing the covariance matrix $\bm\Sigma$ through a covariate-dependent modified Cholesky decomposition. While \citet{KleKneKlaLan2015,MusMaySimUmlZei2022} develop Bayesian inference based on Markov chain Monte Carlo (MCMC) simulations, \citet{GioFasBrowBel2022} implement a gradient boosting approach.  Conceptually, all these approaches extend the 
 famous seemingly unrelated regression model  \citep[SUR;][]{Zel1962}, which combines univariate Gaussian mean regression models with a covariate-independent Gaussian correlation structure. 
However, SUR models and multivariate Gaussian GAMLSS enforce a joint parametric assumption implying that all margins are Gaussian distributions. This assumption is too restrictive in many cases. For example, our application on traffic detection in Berlin requires both discrete and non-Gaussian continuous margins.

In this paper, we present an approach to modelling the distribution of a multivariate response $\bm{Y}=(Y_1,\ldots,Y_D)^\top$ conditional on a set of
covariates $\bm x$ that does not require the response to be jointly Gaussian. Our method complements existing models from the literature and is based on the framework of GAMLSS leveraging Gaussian copula \citep{Son2000} to model the dependence structure between the different response variables $Y_d$, $d=1,\ldots,D$. Unlike the aforementioned contributions in the field, our model however allows for arbitrary parametric non-Gaussian marginal distributions. Each parameter of the resulting multivariate distribution (including those parameterizing the correlation matrix of the Gaussian copula) is linked to a covariate-dependent additive predictor. We use Bayesian inference to estimate the parameters of the model and demonstrate its applicability on simulated and real data examples. The approach is truly multivariate in the sense that it is not restricted to bi- or trivariate distributions. 

In general, copula models overcome the restriction of a joint parametric distribution assumption by modelling the joint dependence structure and the marginal distributions separately; and copula regression has become popular to analyse multivariate regression data \citep[e.g.,][]{CraSab2012,KraBreSilCza2013}. 
For instance, \citet{PitChaKoh2006} use a Gaussian copula to model the dependence between non-Gaussian margins. Within the GAMLSS framework, \citet{MarRad2017,HanKleFasSchMay2022}  develop bivariate one-parameter copula regression models. Conceptually, vector generalized additive models \citep[VGAM;][]{Yee2015} use the same idea as bivariate copula GAMLSS but predictor specifications are more restrictive. \citet{SonLiYua2009,AleBot2021} combine separate one-dimensional GLMs into a multivariate regression model through the Gaussian copula. These approaches can be extended to more flexible non-Gaussian copula models, e.g., using Archimedean  copulas \citep{Nel2006} or vine copulas \citep{CzaNag2022}. However  these models are usually less flexible than our approach, in the sense that the parameters of the copula may not depend on covariates \citep[e.g.,][]{DasPraGon2020,CzaNag2022}, are restricted to the bivariate case \citep[e.g.,][]{OakRit2000, VatCha2015, SunDin2021} or assume marginal distributions from the exponential family \citep{VatNag2018}.


Our work is closely related to multivariate conditional transformation models \citep[MCTMs;][]{KleHotBarKne2022}. MCTMs model a multivariate response through a Gaussian copula with flexible marginal distributions based on univariate conditional transformation models \citep{HotKneBuh2014}. In this way, non-linear effects on all aspects of the marginal distributions as well as on the dependence structure can be captured. However, this approach does not easily translate to additive type predictor models but is so far mostly restricted to models with only one covariate and unpenalized effects, while our Bayesian approach easily allows for data-driven smoothness or variable selection through appropriate prior distributions. 

To model the dependence structure, we employ a computationally attractive modified Cholesky decomposition of the correlation matrix of the Gaussian copula. Similar decompositions have been used successfully in multivariate Gaussian regression and MCTMs, as the entries of the correlation matrix can be estimated unrestrictedly, while enabling a direct relation to independence between specific response components as special cases.  Other notable attempts to covariance modelling include \citet{WuPou2003}, who parameterizes a multivariate Gaussian distribution by imposing an autoregressive structure on the precision matrix, and \citet{BroGilFas2022}, who estimate unrestricted covariate-dependent covariance functions and find the nearest positive definite matrix in a post-processing step. 

Estimation in our complex and high-dimensional model is challenging. For this, we propose efficient Bayesian estimation based on a blockwise Gibbs-within Metropolis-Hastings (MH) Markov chain Monte Carlo (MCMC) algorithm. The MH-steps employ adaptive  iteratively weighted least squares \citep[IWLS;][]{Gam1997} proposal distributions. IWLS are constructed from a second order Taylor approximation to the target posterior
and therefore adjust to the structure of the posterior distribution automatically. This ensures high acceptance rates and therefore efficient sampling. 

Overall, our approach complements alternative methods from the literature and comes with the following four main contributions. (i) Our model is not restricted to two or three dimensional responses. (ii) The marginal distributions can have any parametric forms, thus in particular not requiring joint normality. Some margins may be continuous, others discrete and making parametric assumptions can be useful in some applications and facilitate interpretability of results. (iii) Covariates can be used to model any parameter of the joint distribution and are not restricted exclusively to either marginal distributions or the Gaussian copula parameters. (iv) The Bayesian treatment allows for uncertainty quantification for any quantity of interest derived from the joint distribution.

We showcase these features empirically in our simulations and two real data examples on childhood malnutrition in developing countries and a yet unexplored data set on traffic detection in Berlin, Germany.
The rest of the paper is organized as follows. Section~\ref{sec: multgamlss} introduces our multivariate Gaussian copula GAMLSS and Bayesian estimation through MCMC. Sections~\ref{sec: simulations} and \ref{sec: applications} contain the simulations and real data illustrations.  The paper concludes with a discussion in Section~\ref{sec: conclusion}.

\section{Truly Multivariate Distributional Regression}\label{sec: multgamlss}
We briefly review univariate GAMLSS in Section~\ref{sec: gamlss}, before we extend this model class to a distributional Gaussian copula model for multivariate responses in Sections \ref{sec: our model}--\ref{sec: gaussian copula}. Bayesian inference using MCMC simulations is treated in Section \ref{sec: mcmc}. 

\subsection{Bayesian Univariate GAMLSS}\label{sec: gamlss}
GAMLSS  assume that the distribution of a univariate response $Y$ given the covariates $\bm{x}\in\dsR^q$ has a parametric density $Y\,|\,\bm x\sim p(y\,|\,\bm\theta(\bm x))$ with $K$ distributional parameters $\bm\theta(\bm x)=(\theta_1(\bm x),\ldots,\theta_K(\bm x))^\top$.  Let  $g_k(\cdot)$ be parameter-specific monotonic link functions with inverses $g_k^{-1}(\cdot)$ chosen to ensure potential parameter space restrictions. Each $\theta_k(\bm x)=g_k^{-1}(\eta_k(\bm x))$, $k=1,\dots,K$, is then linked to a structured additive predictor $\eta_k(\bm x)$ of the form 
\begin{equation}\label{eq: additive_predictor}
\eta_k(\bm x)=\beta_{0k}+f_{1k}(\bm x)+\dots+f_{S_kk}(\bm x),
\end{equation}
where $\beta_{0k}$ and $f_{sk}(\bm x)$ are parameter-specific intercept terms and functional effects, respectively.  In general, each $f_{sk}(\bm x)$, $s=1,\ldots,S_k$, $k=1,\ldots,K$, usually depends on different subsets of the entire covariate information $\bm x$ and is approximated through appropriate basis expansions, such as certain spline basis functions to model non-linear effects of continous covariates. This allows us to write $
    f_{sk}(\bm x)=\sum_{l=1}^{L_{sk}}Z_{sk,l}(\bm{x})\beta_{sk,l}=\bm Z_{sk}^\top(\bm x)\bm \beta_{sk},
$
where $\bm Z_{sk}(\bm x)
$ and $\bm \beta_{sk}$ are $L_{sk}$-dimensional column vectors of basis function evaluations and  regression coefficients to be estimated, respectively. To enforce a data-driven amount of smoothness on the functional estimates, the vectors of regression coefficients are typically supplemented with appropriate regularization or penalty terms. In our Bayesian framework we assume  (partially improper) hierarchical and conditionally independent Gaussian priors for  each vector $\bm\beta_{sk}$, that is,
\begin{align}\nonumber
    p(\bm\beta_{sk}\mid\tau^2_{sk})&\propto \left(\tau^2_{sk}\right)^{-\frac{\text{rk}\left(\bm K_{sk}\right)}{2}} \exp\left(-\frac{1}{2\tau^2_{sk}}\bm\beta^\top_{sk}\bm K_{sk}\bm\beta_{sk}\right)\,\mathcal{I}(\bm A_{sk}\bm \beta_{sk}=0),\\
    p(\tau^2_{sk})&\sim \IGD(a_{sk},b_{sk}),\quad s=1,\ldots,S_k,\;k=1,\ldots,K, \label{eq: prior tau}
\end{align}
where $\mathcal{I}(A)$ is one if $A$ is true and zero otherwise. Above, the matrices $\bm K_{sk}$ are prior precision matrices enforcing desired smoothing properties that are specific to $f_{sk}(\bm x)$ and the matrices $\bm A_{sk}$ are chosen to enforce certain constraints on $f_{sk}$. In our case, they are used to overcome the identifiability issues inherent to additive models. We assume constraints of the form $\bm A_{sk}\bm \beta_{sk}=0$ leading to centered effects and priors that are proper conditional on this constraint. A common choice also used throughout this paper is $\bm A_{sk}=(1,\ldots,1)$.    Finally, $\IGD(a,b)$ denotes an inverse gamma distribution with density $\frac{b^a}{\Gamma(a)}\left(\tau^2\right)^{-a-1}\exp\left(-\frac{b}{\tau^2}\right)$ and we set $a_{sk}=0.001=b_{sk}$ as a  default. This hierarchical prior specification for the regression coefficients in \eqref{eq: prior tau} has emerged as a popular default choice for univariate GAMLSS. On  the one hand, it works well empirically and lead to coverages of credible intervals close to the nominal levels when using MCMC.  On the other hand, sufficient conditions for the propriety of the posterior have been derived in this setting \citep{KleKneLan2015}. We adopt  \eqref{eq: prior tau} to our multivariate GAMLSS framework and the theoretical results of \citet{KleKneLan2015} thereby also carry over to our case. 

In principle, it is possible to extend our approach to more complex priors, including effect selection priors and alternative hyperpriors for $\tau_{sk}^2$ such as scale dependent priors \citep{KleKne2016b} and spike and slab priors for effect selection \citep{KleCarKneLanWag2021}.

The general representation of model terms via the design and prior precision matrices $\bm Z_{sk}$ and $\bm K_{sk}$, respectively,  allows to model various effect types \citep[see e.g.,][for a comprehensive overview]{FahKneLanMar2013}. For our simulations and applications, we make the following choices:
\begin{itemize}
    \item a constant term with $\bm Z_{sk}(\bm x)=1$ and flat prior $p(\beta_{0k})\propto 1$ by setting $K_{0k}=0$, 
    \item linear effects, where $\bm Z_{sk}(\bm x)=\bm x_{sk}$ is a vector of original covariates, and   ridge-type priors with $\bm K_{sk}=\bm I$;
    \item non-linear effects  of univariate continuous covariates, where $\bm Z_{sk}(\bm x)$ are cubic B-spline basis functions \citep{EilMar1996} with equidistant knots and   $\bm K_{sk}$ being a second order difference  matrix  \citep{LanBre2004};
    \item random effects $f_{sk}(x_{sk})=\sum_{g=1,\dots,G}\beta_g \mathds{1}(x_{sk}=g)$ for a grouping variable with levels $g\in\{1,\dots,G\}$ with  iid priors, where $\bm K_{sk}=\bm I$; and
    \item discrete spatial effects $f_{sk}(\bm x)=f_{sk}(s_i)$, where $s_i$ denotes the region observation $i$ is located in, such that $\bm Z_{sk}(\bm x)$ is an adjacency matrix   and $\bm K_{sk}$ is a Markov random field prior \citep{RueHel2005}.
\end{itemize}

\subsection{Multivariate Gaussian Copula GAMLSS} \label{sec: our model}
To model the joint distribution of multiple dependent responses of potentially distinct types (such as discrete, categorical, mixed) we combine $D$ univariate GAMLSS with the Gaussian copula as follows. 
Assume  a $D$-dimensional response   $\bm Y=\left(Y_{1},\dots,Y_{D}\right)^\top$ with covariate information $\bm x\in\mathbb{R}^q$. Each marginal distribution is modelled through a  parametric distribution $Y_{d}\mid\bm x\sim F_d\left(y_{d};\bm\theta_{(d)}(\bm x)\right)$ with density $p_d(\cdot\,|\,\bm\theta_{(d)}(\bm x))$ depending on $K_d$ distributional parameters $\bm\theta_{(d)}(\bm x)=\left(\theta_{(d)1}(\bm x),\dots,\theta_{(d)K_d}(\bm x)\right)^\top$ for $d=1,\ldots,D$ as described in Section \ref{sec: gamlss}. 

To construct a joint distribution and to allow for covariate-dependent associations between the components of $\bm Y$, we link the $D$ margins through a Gaussian copula with correlation matrix   $\bm\Omega(\bm x)$. Let   $\bm u=(u_{1},\ldots,u_{D})^\top\sim\ND_D(\bm{0},\bm\Omega(\bm x))$ be multivariate Gaussian distributed with correlation matrix $\bm\Omega(\bm x)$ and zero mean. Then, we assume that for $d=1,\ldots,D$, 
    $y_{d}=F_d^{-1}\left(\Phi_1\left(u_{d}\right); \bm\theta_{(d)}(\bm x)\right)$, 
where $\Phi_1$ is the univariate standard Gaussian cumulative distribution function (CDF). To get a valid copula model, $\bm\Omega(\bm x)$ is restricted to be a correlation matrix and is parameterized through the $\binom{D}{2}=D(D-1)/2$-dimensional column vector  $\bm\lambda(\bm x)=\left(\lambda_{21}(\bm x),\dots,\lambda_{D(D-1)}(\bm x)\right)^\top$ further specified in in \eqref{eq: lambda_matrix} below.  Note that if a marginal CDF $F_{\tilde d}$, $\tilde d\in\lbrace 1,\ldots,D\rbrace$ is non-continuous, $u_{\tilde d}$ is replaced by the randomized quantile residual $\Phi^{-1}\left(a_{\tilde d}+\zeta_{\tilde d}(b_{\tilde d}-a_{\tilde d})\right)$, where $\zeta_{\tilde d}\sim\mathcal{U}(0,1)$ is standard uniformly distributed, $b_{\tilde d}=F_{\tilde d}(y_{\tilde d};\theta_{(\tilde d)}(\bm x))$ and $a_{\tilde d}=\lim\limits_{y\uparrow y_{\tilde d}} F_{\tilde d}(y;\theta_{(\tilde d)}(\bm x))$ are the right- and left-handed limits  of $F_{\tilde d}$ \citep{DunSmy1996,SmiKha2012}.  

Overall, the joint CDF of $\bm Y\mid \bm x$   is given by
\begin{equation*}
    F(\bm y;\bm\theta_{(1)}(\bm x),\ldots,\bm\theta_{(D)}(\bm x),\bm\Omega(\bm x))=\Phi_D\left[\Phi_1^{-1}(F_1(y_1;\bm\theta_{(1)}(\bm x))),\dots,\Phi_1^{-1}(F_d(y_D;\bm\theta_{(D)}(\bm x)); \bm\Omega(\bm x)\right],
\end{equation*}
where $\Phi_D(\cdot; \bm\Sigma)$ denotes the CDF of a $D$-variate Gaussian distribution with zero mean and covariance matrix $\bm\Sigma$. The joint CDF has $K=\sum_{d=1}^DK_d+\binom{D}{2}$ distributional parameters denoted by the stacked vector $\bm\theta(\bm x)=\left(\bm\theta_{(1)}(\bm x)^\top,\dots,\bm\theta_{(D)}(\bm x)^\top,\bm\lambda(\bm x)^\top\right)^\top$. Each element in $\bm\theta(\bm x)$ is related to covariates via predictors of the form \eqref{eq: additive_predictor}.  
\subsection{Parameterization of the Gaussian Copula }\label{sec: gaussian copula}

Motivated by the successful usage in multivariate Gaussian regression and MCTMs, we parameterize $\bm\Omega(\bm x)$ through an unrestricted Cholesky factor of the precision matrix as follows.
Write $\bm\Omega(\bm x)=\diag(\bm\Sigma(\bm x))^{-\frac{1}{2}}\bm\Sigma(\bm x)\diag(\bm\Sigma(\bm x))^{-\frac{1}{2}}$ with covariance matrix $\bm\Sigma(\bm x)=\left(\bm\Lambda(\bm x)\bm\Lambda(\bm x)^\top\right)^{-1}$ and \begin{equation}\label{eq: lambda_matrix}
    \bm\Lambda(\bm x)=
    \begin{pmatrix}
      1 & 0 & \hdots & 0\\
      \lambda_{21}(\bm x)&1&\ddots&\vdots\\
      \vdots & \ddots &\ddots & 0\\
      \lambda_{D1}(\bm x)&\hdots&\lambda_{D(D-1)}(\bm x)&1
    \end{pmatrix}
  \end{equation} 
Note that in contrast to decompositions considered in multivariate Gaussian regression and MCTMs, our parameterization incorporates the normalizing term $\diag(\bm\Sigma(\bm x))^{-\frac{1}{2}}$, since the Gaussian copula is parameterized in terms of the correlation matrix only and not in terms of the covariance matrix $\bm\Sigma(\bm x)$. Using \eqref{eq: lambda_matrix} has the advantage that $\bm\lambda(\bm x)$ is unrestricted and can therefore  directly be linked to additive predictors. However, the entries $\omega_{ld}(\bm x)$, $0\leq l,d\leq D$, of $\bm\Omega(\bm x)$ are not only linked to $\lambda_{ld}(\bm x)$, $1\leq d < l \leq D$, but to a complex non-linear combination of the whole vector $\bm\lambda$, making direct interpretation of the individual parameters challenging. In general, the direct interpretability of each entry in the Cholesky decomposition is lost when moving from the precision matrix $\bm\Sigma(\bm x)^{-1}$ to the correlation matrix $\bm\Omega(\bm x)$ and neither the entries of $\bm\Lambda(\bm x) \bm y$ nor of $\bm\Lambda(\bm x) \bm u$ are necessarily independent. Nevertheless, an interpretation of $\bm\lambda(\bm x)$ is still possible in some common cases.

For instance, in the special case that all margins are Gaussian, $\bm\Omega(\bm x)$ is the correlation matrix of the multivariate Gaussian distribution. Especially, for the bivariate Gaussian case our parameterization is identical (up to the leading sign) to the parameterization of the correlation coefficient in \citet{KleKneKlaLan2015}. \citet{MurDunCarLuc2013} note that if all margins are continuous, zeros in $\bm\Omega(\bm x)^{-1}$ and therefore in $\bm\Lambda(\bm x)\bm\Lambda(\bm x)^\top$ imply conditional independence.

Additionally, if all margins are continuous, the pairwise Kendall's tau $\tau_{ld}(\bm x)$ and Spearman's rho $\rho_{ld}(\bm x)$ quantifying the dependence between $Y_l$ and $Y_d$ are monotonic transformations of the entry $\omega_{ld}(\bm x)$, $0\leq l,d\leq D$ \citep{FanFanKot2002,HulLin2002}.
If at least one of the margins is discrete these direct interpretations no longer hold. However, the influence of covariates on the dependence structure can still be interpreted, for example by investigating mean effect plots as done in Section~\ref{sec: applications}. 

Alternative parameterizations of the Gaussian copula model are discussed in \citet{PitChaKoh2006,MasVar2012,MurDunCarLuc2013}. These  however enforce complex constraints on the components of the parameter vector and can therefore not be linked to independent additive predictors.

\subsection{Likelihood Function}

Assume now, we observe $n$ data pairs $\left\lbrace(\bm y_i,\bm x_i)\right\rbrace_{i=1,\dots,n}$ of conditionally independent $D$-dimensional response vectors $\bm y_{i}=\left(y_{i1},\dots,y_{iD}\right)^\top$ and covariate information $\bm x_i\in\mathbb{R}^q$. Let furthermore $\bm u_{i}=\left(u_{i1},\dots,u_{iD}\right)^\top$ denote the observation-specific vector of (potentially randomized) quantile residuals $u_{id}$, $d=1,\dots,D$. Then, following \citet{Son2000} the joint likelihood is
\begin{align}\label{eq:lik}
\mathcal{L}(\bm\theta(\bm x))=\prod_{i=1}^n\left\{\frac{1}{\sqrt{\det(\bm\Omega(\bm x_i))}}\exp\left(-\frac{1}{2}\left[\bm u_{i}^\top(\bm\Omega(\bm x_i)^{-1}-\bm I_D)\bm u_{i}\right]\right)\prod_{d=1}^D p_d(y_{id}\,|\,\bm\theta_{(d)}(\bm x_i))\right\},
\end{align}
where $\bm I_D$ is a $D\times D$ identity matrix. A thorough derivation of the likelihood in \eqref{eq:lik} is provided in Supporting Information A.

\subsection{Posterior Sampling}\label{sec: mcmc}
Let 
$\bm\beta=(\bm\beta_{01}^\top,\ldots,\bm\beta_{S_1 1}^\top,\ldots,\bm\beta_{0K}^\top,\ldots,\bm\beta_{S_K K}^\top)^\top\in\dsR^{\sum_{k=1}^K\sum_{s=1}^{S_k} L_{sk}+K}$ and\\
$\bm\tau^2=(\tau_{11}^2,\ldots,\tau_{S_1 1}^2,\ldots,\tau_{1K}^2,\ldots,\tau_{S_K K}^2)^\top\in\dsR_{> 0}^{\sum_{k=1}^KS_k}$ denote the stacked vector of regression coefficients and prior variances, respectively.  To sample from the joint posterior $p(\bm\beta,\bm\tau^2\,|\,\bm y)$, we suggest the following  MCMC algorithm. 

\noindent\underline{Step 0} Initialize $\bm \beta$, $\bm \tau^2$.
\\
\noindent For each step of the MCMC sampler, iterate over  predictors $k=1,\ldots,K$ and  effects $s=0,\dots,S_k$ as follows.

\underline{Step 1} Generate from $p(\bm\beta_{sk}|\bm\beta\setminus\bm\beta_{sk},\bm\tau^2,\bm y)$ using  MH steps.

\underline{Step 2} Generate from $p(\tau_{sk}^2|\bm\beta,\bm\tau^2\setminus\tau_{sk}^2,\bm y)$ using Gibbs steps.\\

At \underline{Step 0}, we initialize the sampler for the full parameter vector $\left(\bm\beta,\bm\tau\right)$ at the maximum a posteriori (MAP) estimates obtained via gradient ascent. 

 For generating the regression coefficients $\bm \beta_{sk}$ at \underline{Step 1}, blockwise IWLS proposals for the MH steps are used. They are constructed from a second order Taylor approximation to the target $p(\bm y|\bm\beta,\bm\tau^2)p(\bm\beta|\bm\tau^2)$ 
and therefore adjust to the structure of the posterior distribution automatically.  We use Newton-Raphson-type updates instead of Fisher-scoring updates, which allows us to use the automatic differentiation of the Python package \texttt{pytorch} \citep{PasGroChiChaYanDevLinDesAntLer2017}. A detailed derivation of the proposal distributions is given in  Supporting Information~B.

For generating the smoothing variances at \underline{Step 2}, Gibbs steps are employed based on the conditional posteriors   
  $$  \tau^2_{sk}\mid \bm\beta,\bm\tau^2\setminus\tau_{sk}^2,\bm y \sim \IGD\left(\frac{\text{rk}(\bm K_{sk})}{2}+a_{sk},\frac{1}{2}\bm\beta_{sk}^\top \bm K_{sk}\bm\beta_{sk}+b_{sk}\right).
$$

\section{Empirical Evaluation}\label{sec: simulations}
The aim of this section is to empirically evaluate the accuracy of  our approach labeled \texttt{multgamlss} compared to existing benchmarks from the literature on a bivariate Gaussian  example and a five dimensional non-Gaussian case. Supporting Information~C contains further details on the benchmarks, additional simulation experiments and results not presented in the main text.

\subsection{Related Work and Benchmarks} \label{sec: benchmarks}
We consider the following competing methods summarized in Table~\ref{tab: comp_benchmarks}. 

\begin{itemize}
\item[\texttt{bamlss}]\citet{KleKneKlaLan2015} propose multivariate distributional regression models based on multivariate parametric distributions within the GAMLSS framework. In particular, they consider a bivariate Gaussian response, with two marginal mean parameters and a covariance matrix which is modelled through three parameters, namely two marginal standard deviations $\sigma_1,\sigma_2$ and a correlation parameter $\rho$. Notably the parameterization of the correlation parameter corresponds to the parameterization of our Gaussian copula introduced in Section~\ref{sec: gaussian copula}, when $D=2$ up to the leading sign. That is $\rho=\frac{\eta}{\sqrt{1+\eta^2}}$, where $\eta$ denotes the linear predictor for $\rho$. Estimation is conducted using MCMC. 

\item[\texttt{vgam}] The VGAM approach of \citet{Yee2015} is conceptually similar to \texttt{bamlss} described above. It allows to specify a bivariate Gaussian likelihood in which each of the five distribution parameters (marginal means, marginal standard deviations and correlation parameter) is linked to an addative predictor via link functions. The default parameterization of the correlation parameter $\rho=\frac{\exp(\eta)-1}{\exp(\eta)+1}$ differs from the parameterization used in \texttt{bamlss}. The model is fitted through a penalized maximum likelihood approach based on Fisher scoring and automatic smoothing parameter selection. 

\item[\texttt{mvngam}] The additive covariance matrix model by \citet{GioFasBrowBel2022} extends \texttt{vgam} to a multivariate Gaussian likelihood beyond $D=2$ dimensions. The $D\times D$ dimensional covariance matrix is modelled through a  modified Cholesky decomposition \citep{Pou1999}, that is $\bm\Sigma^{-1}=\bm\Lambda^\top \bm\Delta^{-2}\bm\Lambda$, where $\bm\Delta$ is a diagonal matrix with positive entries and $\Lambda$ is a lower triangular matrix with unit diagonal.  In \texttt{multgamlss} $\bm\Delta$ is restricted to be the identity matrix. \citet{GioFasBrowBel2022} use gradient boosting to fit their model. 

\item[\texttt{mctm}] MCTMs \citep{KleHotBarKne2022} combine $D$ univariate conditional transformation models  by finding a covariate-dependent transformation function $h$, such that $h(\bm y)=(h_1(y_1),\dots,h_D(y_D))$ is standard Gaussian distributed. Instead of linking parameters of the response distribution directly to covariates, MCTMs learn  the covariate-dependent transformation function. The induced dependence structure can be related to a Gaussian copula due to the triangular structure of the joint distribution.  MCTMs are fitted through a maximum likelihood approach using constraint optimization.

\item[\texttt{mle/map}]
It is possible to consider maximum likelihood (labeled \texttt{mle}) or MAP (labeled \texttt{map}) estimation for our multivariate Gaussian copula GAMLSS model. Both can be  estimated through a simple stochastic gradient ascent scheme using automatic differentiation. Confidence sets for both \texttt{map} and \texttt{mle} could be obtained via bootstrap, making them a potential alternative to the proposed MCMC sampler. We include \texttt{mle} and \texttt{map} as additional benchmarks in our simulation study. 
\end{itemize}

\begin{table}[ht]
\centering \renewcommand{\arraystretch}{0.75}
\begin{tabular}{l|ccc}
 & \thead{Dimension of \\ response $D>2$} & \thead{Combination of different \\ marginal distributions} & \thead{Combination of different \\ effect types through \\additive predictors} \\ \hline
\texttt{multgamlss} & \checkmark & \checkmark & \checkmark \\
\texttt{bamlss} &  &  & \checkmark \\
\texttt{vgam} &  &  & \checkmark \\
\texttt{mvngam} & \checkmark &  & \checkmark \\
\texttt{mctm} & \checkmark & \checkmark & 
\end{tabular}
\caption{\small Comparison between the proposed \texttt{multgamlss} approach and considered benchmark methods from the literature.} \label{tab: comp_benchmarks}
\end{table}

\subsection{Bivariate Gaussian Distribution}\label{sec: bivariate_gaussian}

\paragraph{Simulation design} We simulate data from a bivariate Gaussian distribution with correlation parameter  $\rho$ and Gaussian margins $y_d\sim\ND(\mu_d,\sigma_d^2)$ $d=1,2$. We generate $250$ independent data sets with $n=100, 250, 500, 1000$ observations using the additive predictor specifications from \citet{KleKneKlaLan2015} with three covariates. Even though the main focus of our method lies in the estimation of multivariate distributions with non-Gaussian margins, this example works well as a proof of concept simulation study as it allows benchmarking against all methods discussed in Section~\ref{sec: benchmarks} but \texttt{mctm}. A fair comparison with a \texttt{mctm} on this data  is not straightforward since the current implementation of the \textsf{R}-package \texttt{tram{}} does not yet support additive predictors. 

\paragraph{Main results} 
Figure C.5 in the Supporting Information shows the logarithmic average mean squared errors (log-AMSE) of \texttt{multgamlss}, \texttt{bamlss}, \texttt{vgam}, \texttt{mvngam}, \texttt{mle} and \texttt{map} for the five distributional parameters.  In terms of the log-AMSE, \texttt{multgamlss} is competitive with the leading benchmark method \texttt{bamlss}.
\texttt{multgamlss} accurately recovers the true effects across the various sample sizes and the coverage rates of point-wise 95\% credible intervals are close to the nominal levels (see Figure~C.6 in the Supporting Information). As anticipated, larger sample sizes yield improved effect estimates and smaller credible intervals.
Generating $1000$ MCMC samples for this set-up with $n=500$ takes about $11$ minutes on a standard laptop. The proposed MCMC sampler yields good mixing and convergence qualities as assessed through a number of best practice diagnostic checks \citep{GelVehSimMarCarYaoKenGabBurMod2020} further discussed in Supporting Information~C.

\subsection{Five-dimensional non-Gaussian Distribution} \label{sec: design 5d}
\paragraph{Simulation design} \label{subsec: design 5d}
To investigate the performance of \texttt{multgamlss} in a truly multivariate setting with non-Gaussian margins, we reanalyze the  example of \citet{KleHotBarKne2022}. We simulate $250$ data sets with $n=500$ observations and a single covariate $x\sim\mathcal{U}(-0.9,0.9)$ from a five dimensional Gaussian copula model with Dagum margins. The true effects on $\bm\lambda$ are $\lambda_{21}=x^2$, $\lambda_{31}=-x$, $\lambda_{32}=x^3-x$ and  $\lambda_{ld}=0$ for $l>d>3$. The marginal parameters $a_d, b_d$ and $p_d$ of the Dagum margins do not depend on $x$ and are randomly drawn as $\log(a_d), \log(b_d), \log(p_d) \sim \mathcal{U}(-1,2),\, d=1,\dots,5.$ This model includes a total of 25 distributional parameters, of which 10 parameters are associated with $\bm{\Omega}$.  
As \texttt{bamlss} and \texttt{mvngam} are not suitable to analyze a copula model with non-Gaussian margins and \texttt{vgam} currently lacks implementation for the five dimensional Gaussian copula, our comparison of performance is focused solely on \texttt{mctm}, \texttt{map} und \texttt{mle}.  Notably, \texttt{mctm} emerges as the strongest competitor due to its flexibility and compatibility with other approaches, as demonstrated on lower-dimensional data sets \citep{KleHotBarKne2022}. 

\begin{figure}[!ht]
    \centering
    \includegraphics[width=0.95\textwidth,keepaspectratio]{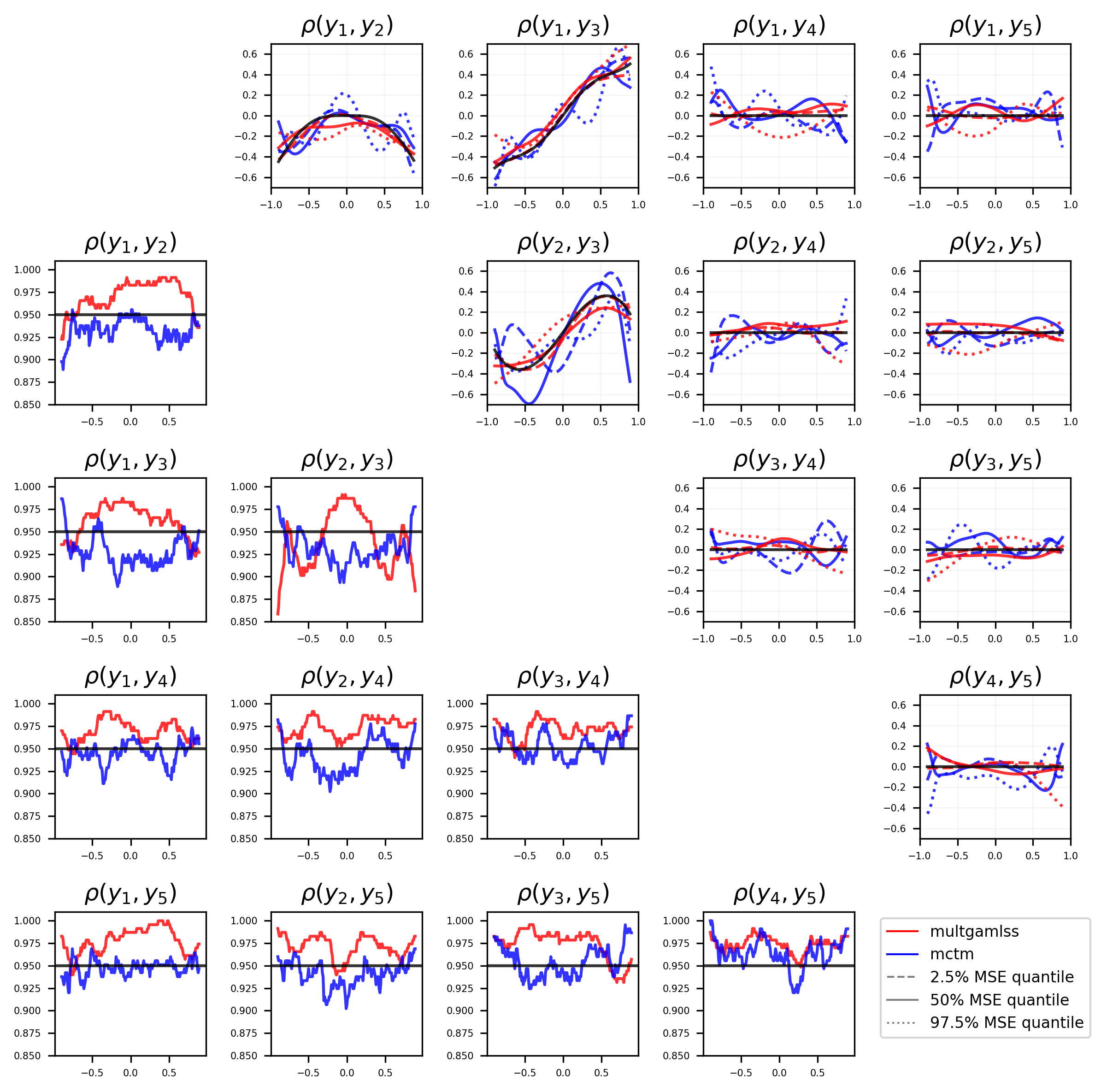}
    \caption{\small Five dimensional non-Gaussian distribution. Upper triangle: Estimated effects on on the pairwise Spearman's rho correlation coefficients. Shown are the $2.5\%$ quantile (dashed), the $50\%$ quantile (bold) and the $95.5\%$ quantile (dotted) quantiles of the MSE-distribution for \texttt{multgamlss} (red) and \texttt{mctm} (blue) across $250$ repetitions. The $x$-axis shows the value of the covariable $x$, while the $y$-axis gives the corresponding correlation coefficient. The black lines correspond to the true effects. Lower triangle: Coverage rates for $95\%$ pointwise Bayesian credible intervals derived through \texttt{multgamlss} (red) and for $95\%$ confidence intervals  derived by \texttt{multgamlss} (blue) are shown in the lower triangular plots. Here, the $y$-axis denotes the observed coverage across the $250$ repetitions.}
    \label{fig: sim high_dims effects}
\end{figure}

\paragraph{Main results}
Since our main interest lies in the recovery of the dependence structure, we focus on the estimated effects on the pairwise Spearman's rho correlation coefficients $\rho(y_k,y_l)=\frac{6}{\pi}\arcsin{\frac{\omega_{kl}}{2}}$,  which are monotonic transformations of the entries $\omega_{kl}$ of $\bm{\Omega}$ for $k,l=1,\dots,5$. The upper triangle of Figure~\ref{fig: sim high_dims effects} summarizes the estimated effects by showing the $2.5\%, 50\%$ and $97.5\%$ quantiles of the mean squared error (MSE) distribution. Both \texttt{multgamlss} and \texttt{mctm} capture the general functional form of the effects well. Especially the true zero effects are identified correctly. \texttt{multgamlss} leads to smoother estimates, while the effects estimated by \texttt{mctm} are slightly wigglier suffering from  outliers at the boundaries, especially when the true effects are zero. This behaviour is expected since in comparison to \texttt{mctm} \texttt{multgamlss} penalizes the wiggliness of the functional effects. Consequently \texttt{multgamlss} yields a lower MSE on all ten pairwise Spearman's rho correlation coefficients compared to \texttt{mctm}, \texttt{map} and \texttt{mle} (Figure C.8 in the Supporting Information). The lower triangle of Figure~\ref{fig: sim high_dims effects} shows the coverage rates for $95\%$ pointwise Bayesian credible sets derived from \texttt{multgamlss} and for $95\%$ bootstrap confidence intervals derived from \texttt{mctm}. The coverage rates are close to the nominal levels for both methods, yet \texttt{multgamlss} tends to overestimate coverages slightly whereas \texttt{mctm} partly underestimates the coverages in the central parts of the covariate space. 
On average the credible intervals estimated by \texttt{multgamlss} are sharper (not shown) with an average  width of $0.31$, compared to the confidence intervals estimated by \texttt{mctm}, with an average band width of $0.44$. Hence, \texttt{multgamlss} yields slightly more efficient uncertainty estimates without requiring an additional bootstrap procedure. 
Generating $1000$ MCMC samples takes about $38$ minutes on a standard laptop. 
\section{Applications} \label{sec: applications}
\subsection{Childhood Malnutrition in Nigeria}
As a first illustration, we consider a data set on childhood malnutrition in Nigeria from 2013 previously analyzed in \citet{KleCarKneLanWag2021,StrKleStaKliMay2023} with data originally taken from the Demographic
and Health Survey (DHS, https://dhsprogram.com). 
While both references use only two of the three available dependent variables (mostly due to model restrictions to bivariate responses), we consider the three variate response vector $(y_\text{stunting},y_\text{wasting},y_\text{underweight})$, where $y_\text{stunting}$, $y_\text{wasting}$ and $y_\text{underweight}$ refer to insufficient height with respect to age,  insufficient weight for height, and insufficient weight for age, respectively.  \citet{KleHotBarKne2022} investigate an older data set on malnutrition in India with a three-variate response vector. However, their model incorporates the age of the child \texttt{cage} as the only covariate, even though a large number of demographic and family-related features are available, see Table E.1 in the Supporting Information of \citet{KleCarKneLanWag2021} for details.

\paragraph{Model specification}
We consider two different models. In the first model, each of the three marginal distributions is assumed to be Gaussian, such that the joint distribution is a multivariate Gaussian distribution with nine predictors $\eta_k, k=1,\dots,9$ referring to the three marginal means $\mu_\text{stunting}, \mu_\text{wasting}$ and $\mu_\text{underweight}$, the three logarithms of the standard deviations $\sigma_\text{stunting}, \sigma_\text{wasting}$ and $\sigma_\text{underweight}$ as well as the three unrestricted parameters $\lambda_1, \lambda_2$ and  $\lambda_3$ of the modified Cholesky decomposition of the correlation matrix. In the second model, each of the marginal distributions follows a Student t-distribution. This model incorporates three additional predictors referring to the logarithmic degrees of freedom $\nu_\text{stunting}, \nu_\text{wasting}$ and $\nu_\text{underweight}$. 
In both models each predictor is that used in \citet{StrKleStaKliMay2023} and is modelled additively through four functional effects of \texttt{cage}, the  age of the child in months, \texttt{mage}, the mother's age and body mass index \texttt{mbmi}, her partner's education \texttt{edupartner}, a discrete spatial effect \texttt{subregion} using the 37 districts in Nigeria. We also include eleven further linear effects of binary and categorical variables, such as \texttt{munemployed}, whether the mother is unemployed and \texttt{csex}, the sex of the child. 

\paragraph{Model fits}
For the Gaussian model all three margins show clear deviations from normality in the right tail (see Figure~D.11 in the Supporting Information). Previous analyses of this data set  did not account for these deviations. An inspection of the normalized quantile residuals suggests that the Student t model fits the data slightly better, so that we consider the Student t model for the rest of this analysis. 

\paragraph{Estimated effects}
The estimated functional and spatial effects coincide well with previous analyzes, see the Supporting Information~D for a discussion of the marginal effects. 

\begin{figure}[ht]
    \centering
    \includegraphics[width=0.95\textwidth,keepaspectratio]{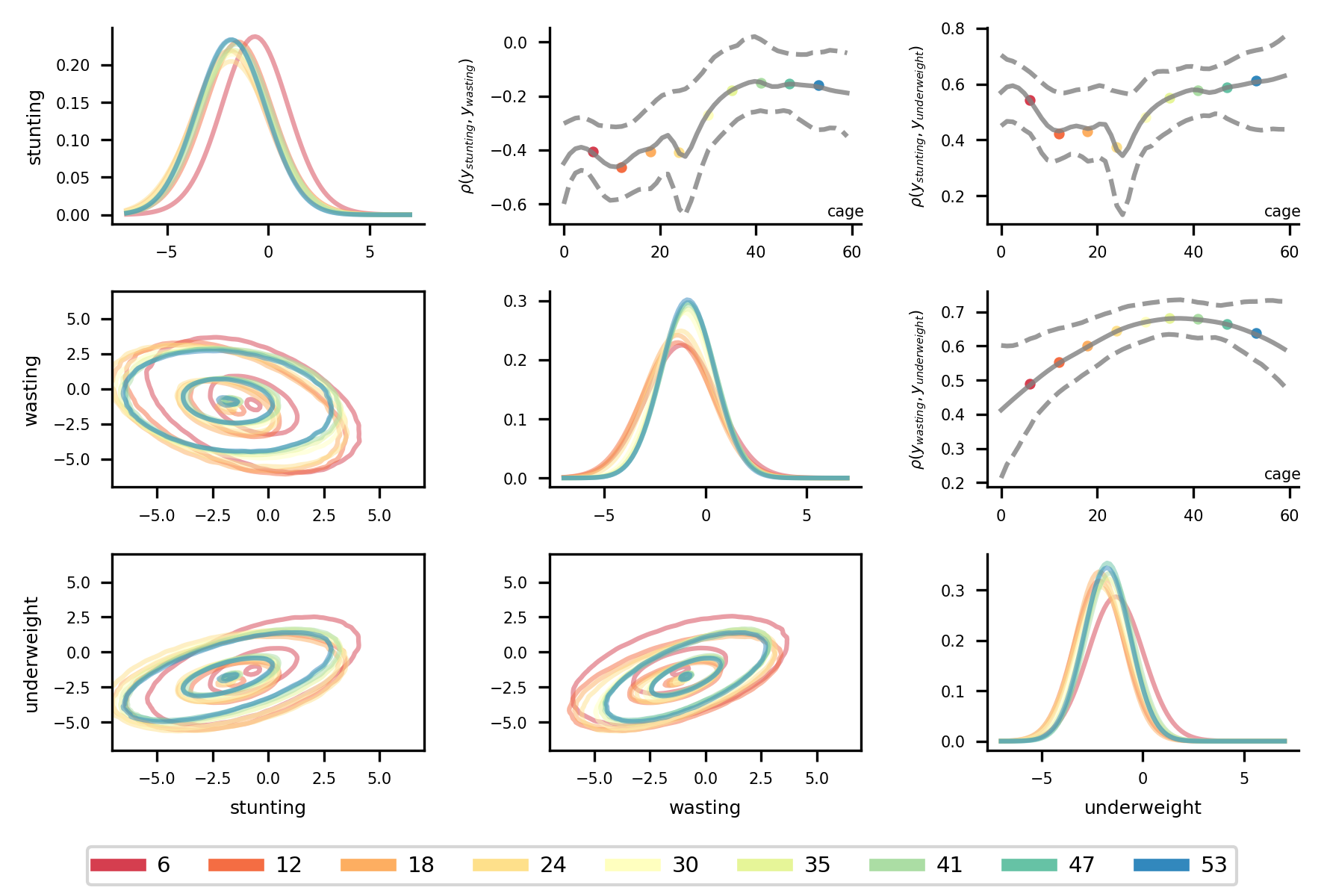}
    \caption{\small Childhood malnutrition. The effects of \texttt{cage} at different ages (indicated by colour). All other covariates are set to fixed values as explained in the main text. On the main diagonal, the marginal densities are summarized, while bivariate contour plots with levels 0.025, 0.5 and 0.975 are given on the lower triangle. The upper triangle shows the estimated effects on the pairwise Spearman's rho correlation coefficient in relation to \texttt{cage}. The posterior mean (bold) and 95\% credible intervals (dashed) are given. The dots indicate the different age levels considered. }
    \label{fig: nigeria_typicallChild}
\end{figure}

The correlation structure is strongly influenced by \texttt{cage}. This becomes apparent when we consider the influence of \texttt{cage} on the joint distribution by setting all other covariates to fixed values (mean for continuous variables and mode for non-continuous variables) as illustrated in Figure~\ref{fig: nigeria_typicallChild}. This figure summarizes univariate marginal densities (diagonal), bivariate margins (lower left) and effects on Spearman's rho (upper right) for increasing \texttt{cage}. We observe that $y_\text{stunting}$ and $y_\text{wasting}$ have a negative Spearman's rho correlation coefficient, which is stronger for children younger than $20$ months. Likewise, $y_\text{wasting}$ and $y_\text{underweight}$ are positively correlated and the correlation increases within the first 20 months. The rank correlation between $y_\text{stunting}$ and $y_\text{underweight}$ is overall high without a strong influence of \texttt{cage}. We do not find a strong effect of either \texttt{edupartner}, \texttt{mage} nor \texttt{mbmi} on the correlation structure. 

\begin{figure}[ht]
    \centering
    \includegraphics[width=0.95\textwidth,keepaspectratio]{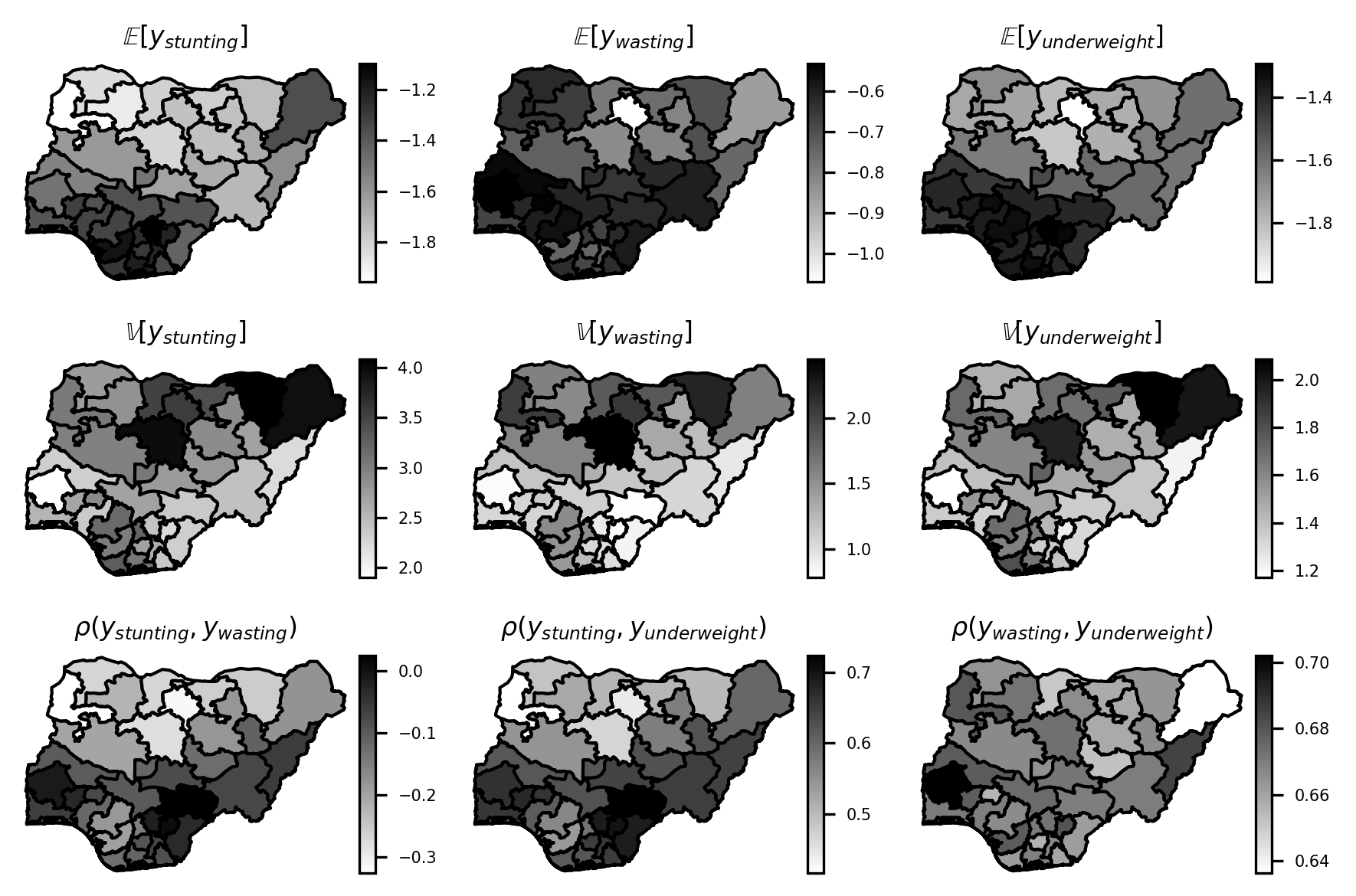}
    \caption{\small Childhood malnutrition. Spatial effects for all $37$ regions. All other covariates are set to fixed values as explained in the main text. The marginal densities are summarized through the effects on the mean (top row) and the variance (middle row). The bottom row shows estimated effects on the pairwise Spearman’s rho correlation coefficient in relation to \texttt{subregion}.}
    \label{fig: nigeria_typicallChild_subregion}
\end{figure}

In Figure~\ref{fig: nigeria_typicallChild_subregion} we vary \texttt{subregion}, while fixing the remaining covariates (mean for continuous variables and mode for binary variables) to investigate the influence of the spatial effect on the joint distribution. The parametric form of the margins allows us to directly consider the influence on $\mathbb{E}\left[y_j\right]=\mu_j$, the expected values, and $\mathbb{V}\left[y_j\right]=\sigma_j^2\frac{\nu_j}{\nu_j-2}$, the variances of the univariate marginal distributions, for $j\in\{\text{stunting, wasting, underweight}\}$. The higher values for $\mathbb{E}\left[y_{\text{stunting}}\right]$ and $\mathbb{E}\left[y_{\text{underweight}}\right]$ in regions in southern Nigeria compared to the average indicate a lower risk for stunted growth and  insufficient weight for age in these regions. Additionally, the variances $\mathbb{V}\left[y_{\text{stunting}}\right]$ and $\mathbb{V}\left[y_{\text{underweight}}\right]$ are lower in southern regions compared to the north of Nigeria. While Spearman's rho correlation coefficient $\rho\left(y_\text{wasting},y_\text{underweight}\right)$ seems to be almost constant with respect to \texttt{subregion}, the spatial effects on $\rho\left(y_\text{stunting},y_\text{wasting}\right)$ and $\rho\left(y_\text{stunting},y_\text{underweight}\right)$ are more pronounced and exhibit a similar structure. 

\subsection{Traffic Detection in Berlin} \label{sec: traffic_detection}
\begin{figure}[ht]
    \centering
    \includegraphics[width=0.95\textwidth,keepaspectratio]{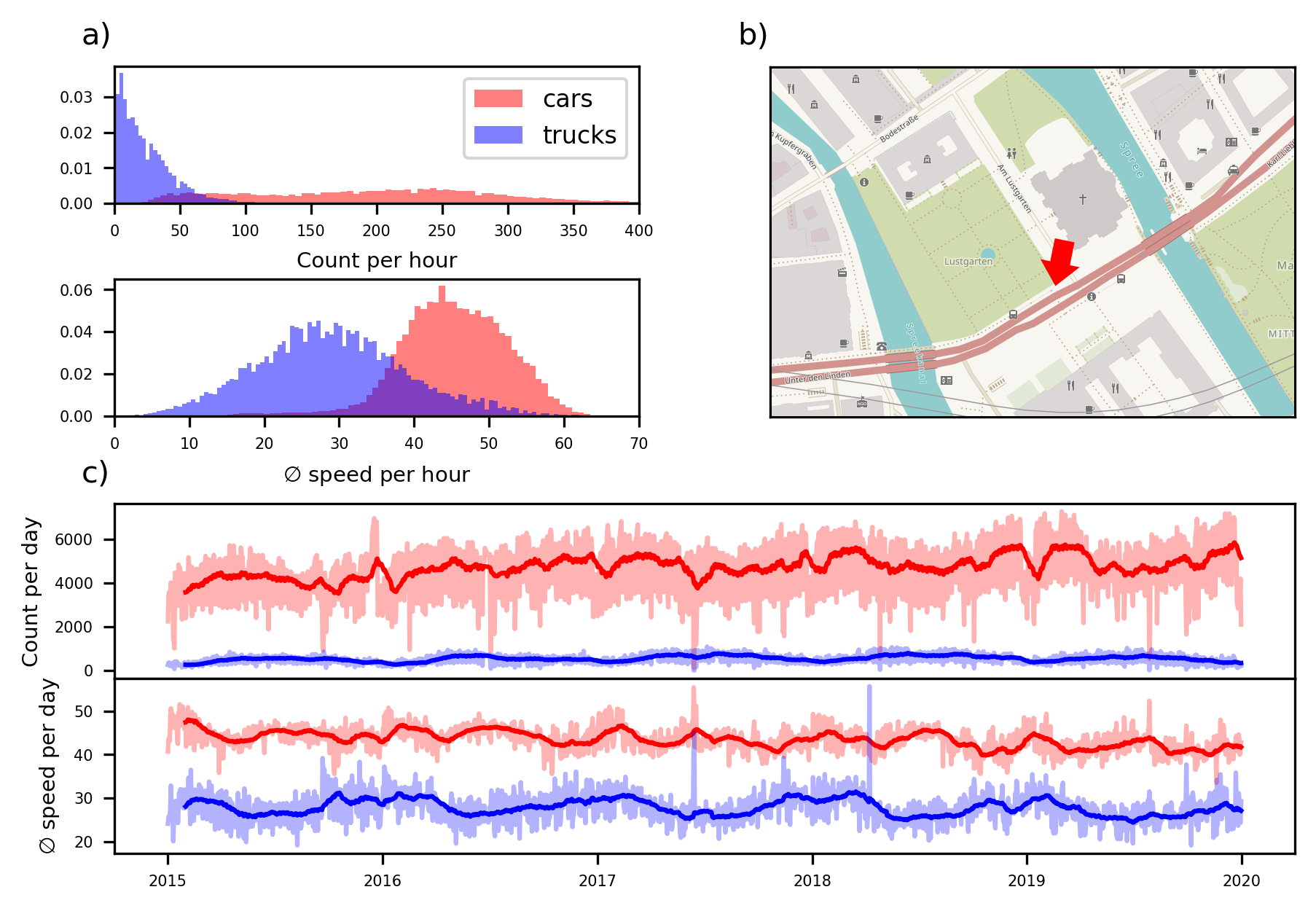}
    \caption{\small Traffic detection. \textbf{a)} Histograms of the counts per hour (top) and the average speed per hour (bottom) of cars (red) and trucks (blue). \textbf{b)} Position of the traffic detection sensor considered. Map data is taken from \textsc{open street map}. \textbf{c)} Time series representation of the count per day (top) and the average speed per day (bottom). The bold line is the 30-day rolling mean.}
    \label{fig: traffic_descriptive}
\end{figure}

In the German capital Berlin, several hundred sensors monitor the traffic along main roads. Hourly aggregated data is publicly available through the Senatsverwaltung für Umwelt, Mobilität, Verbraucher- und Klimaschutz and Verkehrsinformationszentrale Berlin (viz.berlin.de). We consider the measures of a single traffic detector located at ``Unter den Linden'' close to the  main building of the Humboldt-Universit\"at zu Berlin (HU) for our analysis. The response variable is four dimensional and consists of the number of cars $y_{c, car}$ and trucks $y_{c, truck}$ passing the sensor, as well as the average speed of the cars $y_{s, car}$ and $y_{s, truck}$ per hour over five years from 2015--2020 leading to a total of $39,739$ data points. 
Analyzing this four dimensional response $\bm y_t=\left(y_{t;c, car},y_{t;s, car},y_{t;c, truck},y_{t;s, truck}\right)^\top$ is challenging since it incorporates discrete and continuous components with a complex dependence structure. Descriptive summaries  of the data can be found in Figure~\ref{fig: traffic_descriptive}. 

\paragraph{Model specification} To gain insights into the temporal traffic dynamics around the HU main building,  we model   $\bm y_t$ jointly through a Gaussian copula  with negative binomial distributions for the discrete margins $y_{c,car}$ and $y_{c, truck}$ and Student t distributions for the continuous margins $\log\left(y_{s,car}\right)$ and $\log\left(y_{s,truck}\right)$.
Each of the 16  predictors  is  
\begin{align*}
    \eta_k = \sum_{h\in\{(c,car),(s,car),(c,truck),(s,truck)\}}\bm{\widetilde{y_{t;h}}}\bm\beta^{(1)}_{hk}+\delta_t \bm{\widetilde{y_{t;h}}}\bm\beta^{(2)}_{hk}
,\end{align*}
where $\delta_t \in \{1,-1\}$ indicates if time point $t$ is on a weekend and $\bm{\widetilde{y_{t;h}}}$ is the vector of normalized lagged values with lags of $1,2,\ldots, 23$ hours, $1,2,\ldots,6$ days, $1,2,3,4,8,12,\ldots,52$ weeks  to account for the auto-regressive structure in the data. For our analysis, we consider only time points $t$ for which the full vector of lagged values $\bm{\widetilde{y_{t;h}}}$ is available. 

\begin{figure}[ht]
    \centering
    \includegraphics[width=0.95\textwidth,keepaspectratio]{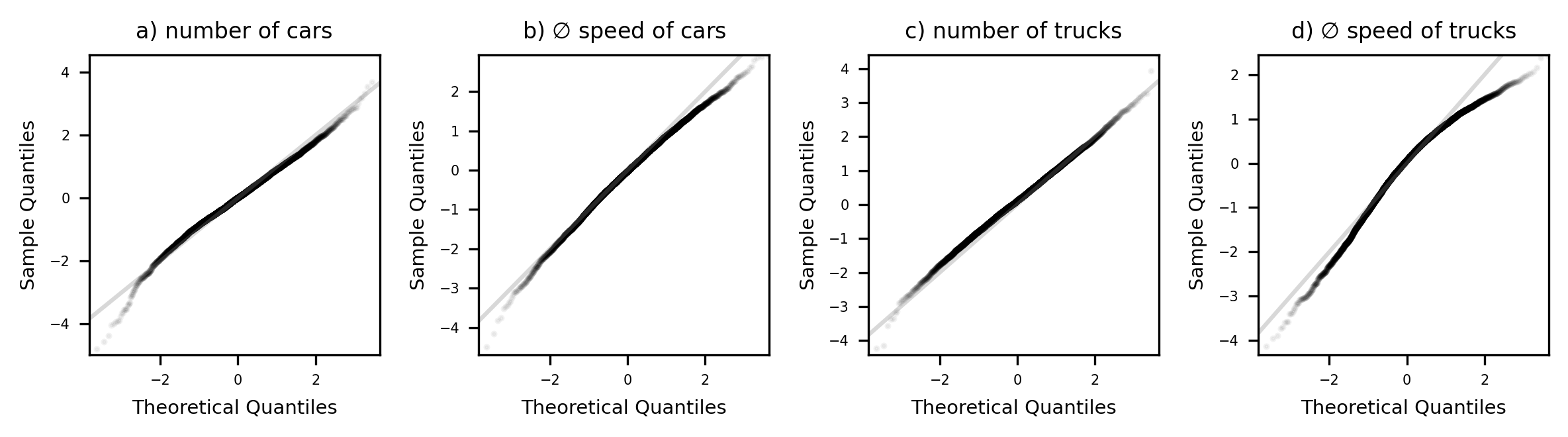}
    \caption{\small Traffic detection. Normalized quantile residuals $\Phi_1^{-1}(F_j(y_{ij};\hat{\bm\theta}(\bm x_i)))$ based on the posterior mean $\hat{\bm\theta}(\bm x_i)$ for $y_{c, car}$, $\log(y_{s,car})$, $y_{c,truck}$ and $\log(y_{s,truck})$ (from left to right).}
    \label{fig: traffic_qq_ts}
\end{figure}

\paragraph{Model fits}
Figure~\ref{fig: traffic_qq_ts} displays normalized quantile residuals for each of the four marginal distributions. The results indicate that the overall fit for the margins is satisfactory. Only, for $y_{(s,truck)}$, there are significant deviations from the distributional assumption in the model, while deviations in the other three marginal distributions are not as pronounced. Minor deviations on the left tail may be attributed to the presence of very small values, which could be due to traffic anomalies, such as line blockage, that are not accounted for in the model. Addressing this issue could involve incorporating additional variables that capture the impact of external factors, such as traffic disruptions or weather conditions. Deviations on the upper tail for $y_{s,car}$ and $y_{s,truck}$ are expected as the Student t margins do not account for the speed of the vehicles being naturally bounded from above.
The copula model is preferable over a model with independent margins according to BIC. 

\paragraph{Estimated effects}

\begin{sidewaysfigure}[htbp]
    \centering
    \includegraphics[width=0.99\textwidth,keepaspectratio]{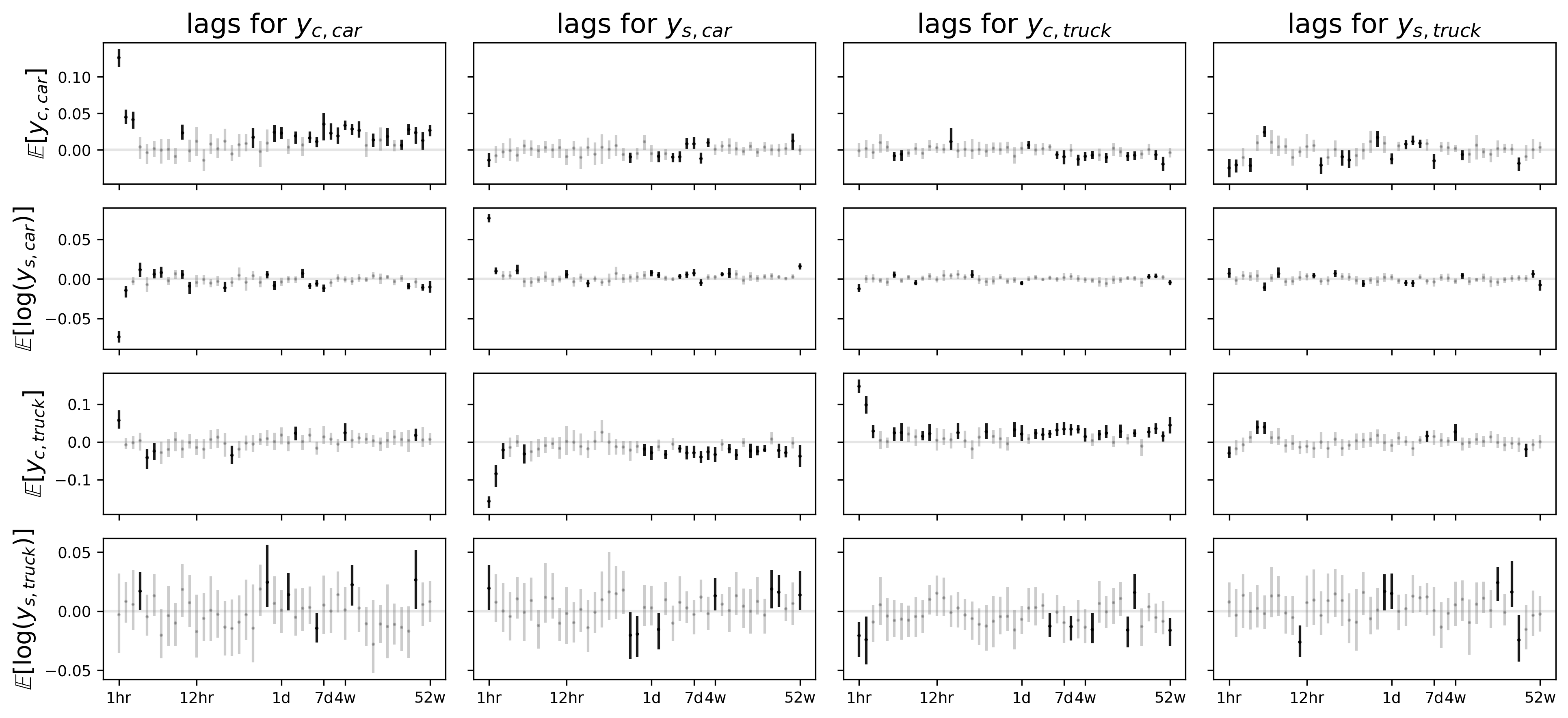}
    \caption{\small Traffic detection. Posterior mean (dots) and posterior $95\%$ credible intervals (vertical lines) for the entries of $\bm\beta^{(1)}_{hk}-\bm\beta^{(2)}_{hk}$, which are the effects for weekdays. If the equally tailed $95\%$ credible interval contains zero, the corresponding effect is grayed. The columns correspond to the standardized lags of $y_{c,car}, y_{s,car}, y_{c,truck}$ and $y_{s,truck}$ respectively. The rows indicate the mean parameter of the four marginal distributions.}
    \label{fig: traffic_effects_mean}
\end{sidewaysfigure}

Figure~\ref{fig: traffic_effects_mean} shows the estimated effects $\bm\beta^{(1)}_{hk}-\bm\beta^{(2)}_{hk}$ on the mean parameters of the four marginal distributions, which are the effects for time points on weekdays. We find that the lagged values of the past few hours have the overall strongest effects. Notably, the marginal distributions are not only influenced by lagged values from their respective marginal, but by the full vector $\bm{\widetilde{y_{t;h}}}$. For example $\mathbb{E}[\log(y_{s,car})]$ is strongly influenced by the number of cars counted in the previous hour. This is reasonable as a high number of cars indicates a busy road resulting in a slower overall speed of vehicles. A description on all fitted effect parameters can be found in Supporting Information~D. 

The mean and  variance of the fitted margins can be calculated for all time points $t$ due to the response distribution being fully parameterized. We find that both marginal count distributions suffer from overdispersion as $\mathbb{V}[y_{t;c,car}]>\mathbb{E}[y_{t;c,car}]$ and $\mathbb{V}[y_{t;c,truck}]>\mathbb{E}[y_{t;c,truck}]$ for all timepoints $t$. The parameter $\omega_{t;3,2}$ controlling the bivariate correlation between $y_{c,truck}$ and $\log\left(y_{s,car}\right)$ is negative for most time points $t$ indicating that a high number of trucks corresponds to the average car driving slower. Similarly, $\omega_{t;4,2}>0$ for most time points $t$ indicating cars and trucks driving faster/slower during similar time periods. 

\begin{figure}[ht]
    \centering
    \includegraphics[width=0.95\textwidth,keepaspectratio]{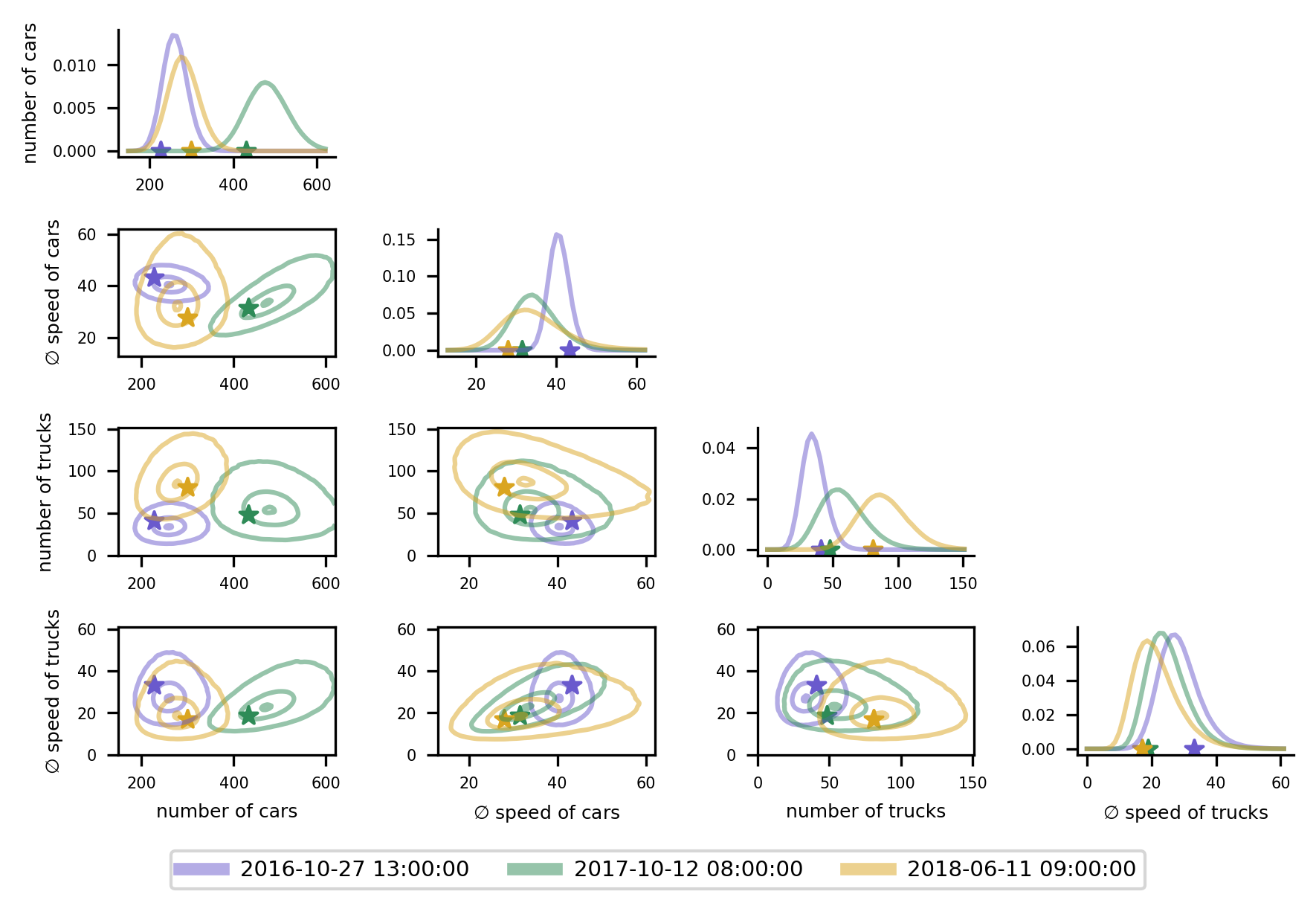}
    \caption{\small Traffic detection. Univariate marginal densities (diagonal) and bivariate contour plots with levels 0.025, 0.5 and 0.975 (lower triangle) for three different time points (indicated by color). The margins are estimated through a kernel density estimator based on a large sample from the fitted distribution. The observed values at the respective time points are given by stars.}
    \label{fig: traffic_fitted_distributions}
\end{figure}

Figure~\ref{fig: traffic_fitted_distributions} summarizes univariate marginal densities (diagonal) and bivariate margins (lower left) for three different, randomly selected time points showcasing the high flexibility of the fitted model. The bivariate contour plots illustrate the influence of the Gaussian copula and correspond well with our findings described above. For example, we find that $y_{s,car}$ and $y_{s,truck}$ are positively correlated for all three time points. 

\section{Summary and Discussion}\label{sec: conclusion}
This paper introduces a  multivariate distributional regression approach  based on a Gaussian copula to model the dependence structure between multiple responses. The proposed model allows for flexible and independent selection of arbitrary parametric marginal distributions for each response component similar to univariate GAMLSS, while incorporating covariate-dependent additive predictors for each distributional parameter (including those parameterizing the correlation matrix of the Gaussian copula). Our model complements existing proposals in the field. Our Bayesian treatment has the appeal to allow for uncertainty quantification, while not being restricted to bivariate responses as many other existing approaches. In our simulation we find comparable performance in situations where competitors are available yet with increased flexibility and modelling options that we showcase on two real data illustrations.

Although the parametrization of the correlation matrix  enables a computationally efficient MCMC sampler, a direct interpretation of additive effects can be challenging. To address this issue, we propose the use of slice plots as a visual inspection tool for the fitted model. However, future research could potentially explore the incorporation of variable and effect selection priors \citep{KleCarKneLanWag2021} into the model to improve interpretability and simplify model selection. 
In this regard,  many alternative shrinkage priors have been considered within the literature \citep[see e.g.,~][]{PowRaf2022}. For instance, incorporating  shrinkage priors that shrink towards independent response components is  a fruitful direction for future research. 

Even though our model formulation is valid for arbitrary dimensions of the response, the proposed MCMC sampler does not scale to arbitrary dimensions due to computational limitations. One reason for this is the high number of parameters involved in the flexible parameterization of the correlation matrix. Future research could investigate alternative parameterizations, which restrict the number of parameters as for instance done in \citet{MusMaySimUmlZei2022}. Alternatively approximate Bayesian inference using variational Bayes \citep{BleKucMca2017} could be better scalable.

\FloatBarrier
\bibliography{bib}
\end{document}